\documentclass{iau}
\usepackage{graphicx}
\usepackage{soul,color,natbib}

\title[Planetary protection in the extreme environments of low-mass stars] 
{Planetary protection in the extreme environments of low-mass stars}

\author[A.~A.~Vidotto]   
{A.~A.~Vidotto$^1$, M.~Jardine$^1$, J.~Morin$^2$, J.-F. Donati$^3$, P.~Lang$^1$ \and A.~J.~B.~Russell$^4$}

\affiliation{$^1$SUPA, University of St Andrews, North Haugh, KY16 9SS, UK \\email: {\tt Aline.Vidotto@st-andrews.ac.uk}\\[\affilskip]
$^2$ Georg-August-Universit\"at, Friedrich-Hund-Platz 1, D-37077, Goettingen, Germany\\[\affilskip]
$^3$ Observatoire Midi-Pir\'en\'ees, 14 Av.~E.~Belin, F-31400, Toulouse, France\\[\affilskip]
$^4$ SUPA, University of Glasgow, University Avenue, G12 8QQ, Glasgow, UK}

\pubyear{2013}
\volume{302}  
\pagerange{119--126}
\jname{Magnetic fields throughout stellar evolution}
\editors{P. Petit, eds.}

\begin{document}
\maketitle

\begin{abstract}
Recent results showed that the magnetic field of M-dwarf (dM) stars, currently the main targets in searches for terrestrial planets, is very different from the solar one, both in topology as well as in intensity. In particular, the magnetised environment surrounding a planet orbiting in the habitable zone (HZ) of dM stars can differ substantially to the one encountered around the Earth. These extreme magnetic fields can compress planetary magnetospheres to such an extent that a significant fraction of the planet's atmosphere may be exposed to erosion by the stellar wind. Using observed surface magnetic maps for a sample of 15 dM stars, we investigate the minimum degree of planetary magnetospheric compression caused by the intense stellar magnetic fields. We show that hypothetical Earth-like planets with similar terrestrial magnetisation ($\sim$1 G) orbiting at the inner (outer) edge of the HZ of these stars would present magnetospheres that extend at most up to 6.1 (11.7) planetary radii. To be able to sustain an Earth-sized magnetosphere, the terrestrial planet would either need to orbit significantly farther out than the traditional limits of the HZ; or else, if it were orbiting within the life-bearing region, it would require a minimum magnetic field ranging from a few G to up to a few thousand G.
\keywords{stars: magnetic fields, stars: planetary systems, stars: rotation, astrobiology}
\end{abstract}
%

Due to  technologies currently adopted in exoplanet searches, dM stars have been the main targets in searches for terrestrial planets. For these stars, the orbital region where a planet should be able to retain liquid water at its surface (known as the habitable zone, HZ) is located significantly closer than the HZ of solar-type stars \citep{1993Icar..101..108K}. These factors make dM stars the prime targets for detecting terrestrial planets in the potentially life-bearing region around the star.  

However, in addition to the retention of liquid water, other factors may be important in assessing the potential for a planet to harbour life. For example, the presence of a relatively strong planetary magnetic field is very likely to play a significant role in planetary habitability. A relatively extended planetary magnetosphere can deflect the stellar wind and other ejecta, protecting the planetary atmosphere against erosion. 

In steady state, the extent of a planet's magnetosphere is determined by force balance at the boundary between the stellar coronal plasma and the planetary plasma. For the planets in the solar system, this is often reduced to a pressure balance at the dayside, the most significant contribution to the external (stellar) wind pressure being the solar wind ram pressure. However, for planets orbiting stars that are significantly more magnetised than the Sun or/and are located at close distances, the stellar magnetic pressure may play an important role in setting the magnetospheric limits \citep{2004ApJ...602L..53I,2009A&A...505..339L,2009ApJ...703.1734V,2010ApJ...720.1262V,2012MNRAS.423.3285V,2011MNRAS.412..351V}. 

In the case of dM stars, their magnetic fields is very different from the solar one (in topology and intensity). Therefore, the magnetised environment surrounding a planet orbiting in the HZ of dM stars can differ substantially to the one encountered around the Earth. In the present work, we quantitatively evaluate the sizes of planetary magnetospheres resulting from the pressure exerted by the intense stellar magnetic fields found around dM stars. Our approach only invokes a stellar magnetic field, neglecting effects such as dynamic pressures. Figure 1a shows the minimum degree of planetary magnetospheric compression caused by the intense stellar magnetic fields for a sample of 15 dM stars whose magnetic fields have been observationally reconstructed \citet{2008MNRAS.390..545D} and \citet{2008MNRAS.390..567M,2010MNRAS.407.2269M}. Hypothetical Earth-like planets with similar terrestrial magnetisation orbiting at the inner (outer) edge of the HZ of these stars would present magnetospheres that extend at most up to $6.1$ (11.7) planetary radii ($r_p$). To be able to sustain an Earth-sized magnetospheres ($\sim 12~r_p$), such planets would require a minimum magnetic field ranging from a few G to up to a few thousand G. Figure 1b shows the closest orbital distance at which an Earth-like planet orbiting the stars in our sample would be able to sustain the present-day Earth's magnetospheric size, assuming it has the same magnetic field as the Earth. Planets orbiting at a closer orbital radius would experience a stronger stellar magnetic pressure, which could reduce the size of the planet's magnetosphere significantly, exposing the planet's atmosphere to erosion by the stellar wind.

\begin{figure}
\centering
	\includegraphics[width=0.95\textwidth]{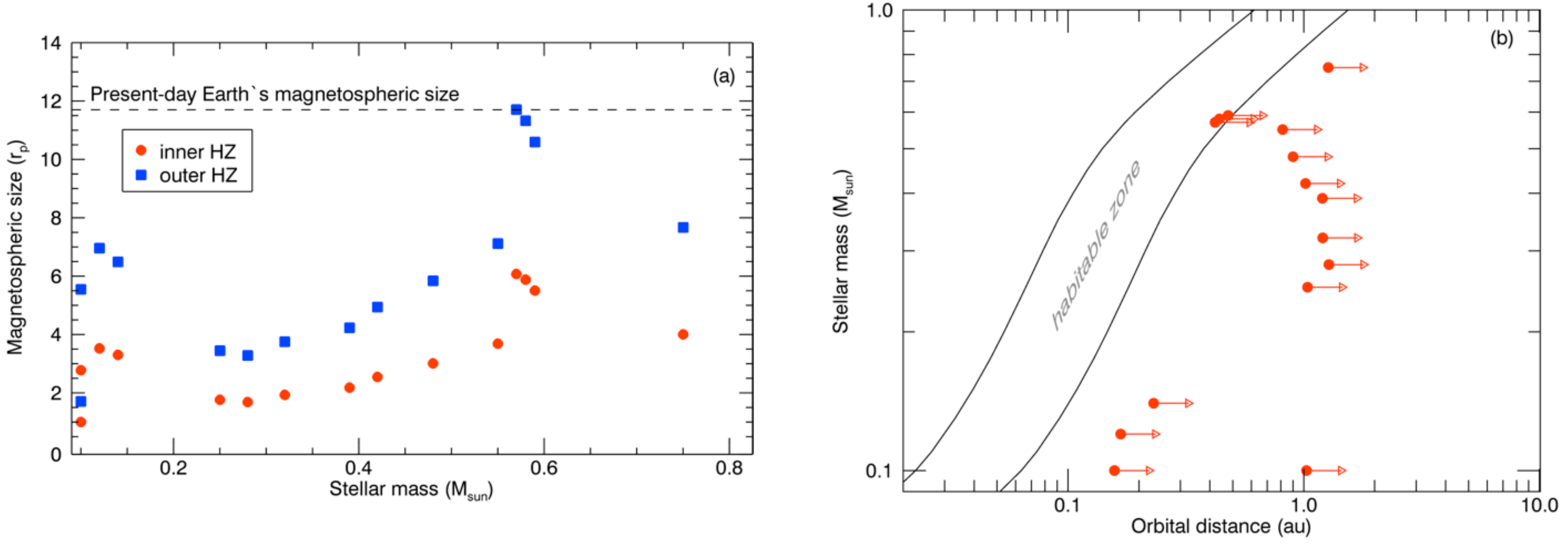}
\caption{(a) The minimum degree of planetary magnetospheric compression caused by the intense stellar magnetic fields. (b) Closest orbital distance at which an Earth-like planet orbiting the stars in our sample would be able to sustain the present-day Earth's magnetospheric size, assuming it has the same magnetic field as the Earth. Adapted from \cite{2013A&A...557A..67V}.}\label{fig.sketch2}
\end{figure}

\bibpunct{(}{)}{,}{a}{}{;} 

\def\aj{{AJ}}                   
\def\araa{{ARA\&A}}             
\def\apj{{ApJ}}                 
\def\apjl{{ApJ}}                
\def\apjs{{ApJS}}               
\def\ao{{Appl.~Opt.}}           
\def\apss{{Ap\&SS}}             
\def\aap{{A\&A}}                
\def\aapr{{A\&A~Rev.}}          
\def\aaps{{A\&AS}}              
\def\azh{{AZh}}                 
\def\baas{{BAAS}}               
\def\jrasc{{JRASC}}             
\def\memras{{MmRAS}}            
\def\mnras{{MNRAS}}             
\def\pra{{Phys.~Rev.~A}}        
\def\prb{{Phys.~Rev.~B}}        
\def\prc{{Phys.~Rev.~C}}        
\def\prd{{Phys.~Rev.~D}}        
\def\pre{{Phys.~Rev.~E}}        
\def\prl{{Phys.~Rev.~Lett.}}    
\def\pasp{{PASP}}               
\def\pasj{{PASJ}}               
\def\qjras{{QJRAS}}             
\def\skytel{{S\&T}}             
\def\solphys{{Sol.~Phys.}}      
\def\sovast{{Soviet~Ast.}}      
\def\ssr{{Space~Sci.~Rev.}}     
\def\zap{{ZAp}}                 
\def\nat{{Nature}}              
\def\iaucirc{{IAU~Circ.}}       
\def\aplett{{Astrophys.~Lett.}} 
\def\apspr{{Astrophys.~Space~Phys.~Res.}}   
\def\bain{{Bull.~Astron.~Inst.~Netherlands}}    
\def\fcp{{Fund.~Cosmic~Phys.}}  
\def\gca{{Geochim.~Cosmochim.~Acta}}        
\def\grl{{Geophys.~Res.~Lett.}} 
\def\jcp{{J.~Chem.~Phys.}}      
\def\jgr{{J.~Geophys.~Res.}}    
\def\jqsrt{{J.~Quant.~Spec.~Radiat.~Transf.}}   
\def\memsai{{Mem.~Soc.~Astron.~Italiana}}   
\def\nphysa{{Nucl.~Phys.~A}}    
\def\physrep{{Phys.~Rep.}}      
\def\physscr{{Phys.~Scr}}       
\def\planss{{Planet.~Space~Sci.}}           
\def\procspie{{Proc.~SPIE}}     

\let\astap=\aap
\let\apjlett=\apjl
\let\apjsupp=\apjs
\let\applopt=\ao
\let\mnrasl=\mnras

\end{document}